\documentclass{fcs}
\usepackage{bm}

\usepackage{multirow}
\usepackage{graphicx}
\usepackage{subfigure}
\usepackage{bm}
\usepackage{diagbox}
\usepackage{enumitem}
\usepackage{url}
\usepackage{caption}
\usepackage{color}
\usepackage{footmisc}
\usepackage{booktabs} 
\usepackage{amsmath}
\usepackage{hyperref}
\usepackage{picins}
\volumn{ }
\doi{ }
\articletype{RESEARCH~ARTICLE}
\copynote{{\copyright} Higher Education Press 2022}
\ratime{Received month dd, yyyy; accepted month dd, yyyy}
\email{{panweike@szu.edu.cn}}

\title{BMLP: Behavior-aware MLP for Heterogeneous Sequential Recommendation}
\author{Wei-Xin~Li$^{1}$, Yu-Hao~Wu$^{1}$, Yang~Liu$^{1}$, Wei-Ke~Pan\xff $^{1}$, Zhong~Ming $^{1,2}$}
\address{	
	{1\quad College of Computer Science and Software Engineering, Shenzhen University, Shenzhen 518060, China
}\\ 
    {2\quad Guangdong Laboratory of Artificial Intelligence and Digital Economy (SZ), Shenzhen 518123, China}
}

\begin{document}
\maketitle
\setcounter{page}{1}
\setlength{\baselineskip}{14pt}

\begin{abstract}
In real recommendation scenarios, users often have different types of behaviors, such as clicking and buying. 
Existing research methods show that it is possible to capture the heterogeneous interests of users through different types of behaviors. 
However, most multi-behavior approaches have limitations in learning the relationship between different behaviors.
In this paper, we propose a novel multilayer perceptron (MLP)-based heterogeneous sequential recommendation method, namely behavior-aware multilayer perceptron (BMLP). 
Specifically, it has two main modules, including a heterogeneous interest perception (HIP) module, which models behaviors at multiple granularities through behavior types and transition relationships, and a purchase intent perception (PIP) module, which adaptively fuses subsequences of {auxiliary} behaviors to capture users' purchase intent. 
Compared {with} mainstream sequence models, MLP is competitive in terms of accuracy {and has} unique advantages in simplicity and efficiency. 
Extensive experiments {show} that BMLP achieves significant improvement over state-of-the-art algorithms on four public datasets. 
In addition, its pure MLP architecture leads to a linear time complexity.
\end{abstract}


\section{Introduction}

\noindent {Recommender systems can effectively alleviate the problem of information overload, particularly with the substantial growth of online users, items, and information in recent years.}
{However,} traditional recommendation methods cannot capture users’  {dynamic} interests and intent.
Sequential recommendation methods, which {aim} to model the behavior sequence of users to obtain more accurate, dynamic, and personalized recommendation lists, overcome this problem well.
\\
\indent Recently, deep learning-based sequential recommendation methods mine users' latent representations and complex sequential relationships from the interaction data. 
RNN is proposed to capture the sequential information between items~\cite{GRU4Rec,RNN_Rec1}. 
CNN-based methods~\cite{Caser, 2019CNN} use filters to learn users' short-term interests. 
After that, attention networks became popular in many fields,  {e.g., being} used to learn the relevant weights between items~\cite{SASRec,2019transformerbased, bert4rec}. 
GNN-based methods~\cite{SRGNN,GNN_Rec1,GNN_Rec2,GNN_Rec3} structure sessions into graphs to capture richer relevance of items. 
Those mainstream deep learning-based methods have achieved exciting results, gradually dominating the recommender systems.
However, they have a limitation {in common}, i.e., they can only deal with {a single} type of behavior. 
\\ 
\indent In a real scenario, the behaviors of users are often rich and diverse, which is unreasonable to regard them as the same behavior type or only select the behaviors of one single type such as purchase to learn the users' preferences. 
There are two categories of works on modeling multiple behaviors. 
Heterogeneous recommendations~\cite{MBGMN, MBGCN, vae++} aggregate representations of various behaviors as overall user preferences. 
They ignore the dynamic dependencies between items and behaviors. 
Heterogeneous sequential recommendation is an emerging and {significant research problem that has relatively few works in its field.}
Specifically, RNN-based methods~\cite{RIB, BINN, RLBL} distinguish the feedback of different behaviors by introducing {behavior} types or integrating {the behavior} information into the RNN module. 
The Transformer-based method~\cite{DMT} obtains multiple interests of the users by modeling the subsequences {of} the same type separately. 
The GNN-based method~\cite{MGNN} captures relationships between the {behaviors} of the same type by constructing a global graph. 
These approaches may ignore complex behavior dependencies and transition relationships. 
Meanwhile, in e-commerce websites, the purchase behavior mainly depends on the overall user preferences and some recent browsing or favoriting behaviors. 
However, for the aforementioned heterogeneous recommendation methods (e.g., MBGMN~\cite{MBGMN}, MBGCN~\cite{MBGCN}), they fuse multiple behaviors to obtain {the} static interest of users, which only considers the overall user preferences. 
{The previous works} for the heterogeneous sequential recommendation can be broadly divided into two branches.
{One} is to model the entire heterogeneous sequence while introducing the behavior information~\cite{RLBL, RIB, BINN}, which may not take adequate advantage of recent {auxiliary} behaviors to capture the intent of the users.
{The other} is to disrupt the integrity of the sequences (e.g., MGNN~\cite{MGNN}, DMT~\cite{DMT}, SDM~\cite{sdm}), modeling the single behavior subsequences separately, which may have a bias in capturing the overall interest of the users. 
Therefore, {we consider some specific challenges of} heterogeneous sequential recommendation: 
(i) The complexity of heterogeneous sequential recommendation models. 
Almost all the works utilize RNN, Transformer {or} GNN to capture sequential {patterns}. 
(ii) The complex dependencies of {behavior information.} 
Distinguishing the feedback of different behaviors and the dependencies between behaviors is critical 
{for accurately learning a user's preferences.}
(iii) The uncertainty of a user's intents. 
In a heterogeneous sequence of intertwined target and {auxiliary} behaviors, it is difficult to accurately predict what the user is likely to purchase. 
\\
\indent To address the above three challenges, we propose a novel heterogeneous sequential recommendation method, i.e., behavior-aware MLP (BMLP), consisting of {a} heterogeneous interest perception (HIP) module and {a} purchase intent perception (PIP) module. 
(i) We use a pure MLP architecture, which has a lower time complexity. 
(ii) {The} heterogeneous interest perception (HIP) module performs multi-granularity processing of the entire heterogeneous {sequence} by introducing behavior types and behavior transition relations. 
It captures the dependencies between behaviors {more comprehensively}. 
(iii) {The} purchase intent perception (PIP) module adaptively fuses the {auxiliary} behaviors such as clicks. 
It captures the potential purchase intent of users more precisely. 
We then conduct extensive experiments on four datasets and find that our BMLP {can}  beat the current mainstream state-of-the-art baselines. 
{We summarize the main contributions of this work as follows.} 

\begin{itemize}[leftmargin=*]
	\item{We propose a novel pure MLP-based recommendation approach {that} is simple and efficient. To the best of our knowledge, this is the first work using MLP to tackle heterogeneous sequential {recommendation}.}
	
	\item { Our HIP module captures the interaction patterns between behaviors from a multi-granularity perspective, namely behavior types and behavior transition relations. Meanwhile, our PIP module aggregates multiple auxiliary behavior subsequences. These two well-designed modules fit well with a user's purchasing habits.}
	
	\item{Compared with the existing state-of-the-art baselines, our BMLP achieves significant improvement on four public datasets.}
\end{itemize}

\section{Related Work}
\indent In this section, we review and summarize some related works on factorization-based and deep learning-based {recommendation}.
\subsection{Factorization-based Recommendation}
\indent In the early studies on sequential recommendation, {FPMC~\cite{FPMC} integrates the  sequential information for personalization and} 
uses Markov chains to capture the first-order relationships between items.
Because the first-order relationship is too simplified, Fossil~\cite{Fossil} extends it to a multi-order dependency. 
Later, TransRec~\cite{TransRec} embeds items as points in a translation space and models users as translation vectors existing in that space.

\indent BF~\cite{BF} uses matrix factorization techniques to model each behavior separately, decomposing users' interest profiles into multiple behavior profiles. 
Based on TransRec~\cite{TransRec}, TransRec++~\cite{Transrec++} introduces behavior transition vectors to further characterize the dependencies between behaviors.

\subsection{Deep Learning-based Recommendation}
\indent \textbf{Sequential Recommendation}. 
In recent years, deep learning has rapidly become popular in sequential recommendation due to its strong modeling ability and generalization. 
With the introduction of deep learning models, the performance has been further improved.  
The recurrent neural network (RNN) based method GRU4Rec~\cite{GRU4Rec} consumes the sequential information and captures the user's dynamic intent very well. 
The convolutional neural network (CNN) based method Caser~\cite{Caser} utilizes some horizontal and vertical filters to search for local information in sequences and global representations of users to record the long-term interests. 
The attention-based method SASRec~\cite{SASRec} emphasizes the important items and downplays the irrelevant items by the attention mechanism. 
The graph neural network (GNN) based method SRGNN~\cite{SRGNN} captures complex transition relationships between items by constructing a graph from a sequence. 
Recently, with the advances in MLP architecture~\cite{MLP-Mixer}, MLP-based recommendation models are springing up. 
MOI-Mixer~\cite{MOI-Mixer} applies this model to sequential recommendation for the first time. 
FMLP~\cite{2022FMLP} introduces a filter layer based on the MLP architecture. 
MLP4Rec~\cite{MLP4Rec} expands the input tensor into 3-D by introducing the item properties.

{
\noindent \textbf{Heterogeneous Non-sequential Recommendation}. 
Many deep learning techniques (e.g., MLP, autoencoder, and GNNs) have been widely adopted in multi-behavior recommendation and have achieved remarkable performance.
EHCF\cite{EHCF} captures the complex relations between different behaviors by relating the transition of each behavior.
VAE++ \cite{vae++}  is a VAE-based method which designs two representation enhancement modules to capture multiple behavior signals.
For the GNN-based methods in multi-behavior recommendation,  MBGCN \cite{MBGCN}  designs a  unified graph for representing multi-behavior data and performs graph convolution operation to learn node representations under different behaviors.
MBGMN \cite{MBGMN}   proposes a meta-graph neural network to capture the complicated dependencies across different types of user-item interactions with the meta-learning paradigm.
CRGCN \cite{CRGCN}  adopts a cascading GCN structure to learn users' preferences under each behavior.
To alleviate the data sparsity and popularity bias, there are some recent works trying to leverage contrastive learning.
CML \cite{CML}  designs a multi-behavior contrastive learning paradigm to capture relations across different behaviors.
MMCLR \cite{MMCLR}  proposes a multi-behavior contrastive learning task and a multi-view contrastive learning task.
MixMBR \cite{MixMBR}  introduces a mixup data augmentation method and combines it with contrastive learning. These methods do not introduce sequential information and differ significantly from our problem definition, for which reason we do not include them in the empirical studies.}

\noindent \textbf{Heterogeneous Sequential Recommendation}. 
Existing heterogeneous sequential recommendation algorithms are still relatively few.  
RLBL~\cite{RLBL} uses {the} behavior transition matrix to represent the relationship between behaviors and uses RNN and log-bilinear to capture a user's interests. 
RIB~\cite{RIB} also introduces the behavior types and incorporates them into the GRU module. 
BINN~\cite{BINN} introduces the behavior information inside LSTM~\cite{LSTM} and learns both the long-term and short-term interests using {users'} past behaviors and items. 
DMT~\cite{DMT} uses Transformer to model the subsequences of each same type of behavior, { which captures the interactions between the same behaviors separately}. 
MGNN-SPred~\cite{MGNN} constructs heterogeneous graphs about multiple types of behaviors. 
MSR~\cite{MSR} uses GNN to model item sequences and GRU to model behavior sequences. 
We can see that these methods are based on RNN, GNN, and Transformer. 
Different from the above methods, we propose a heterogeneous sequential recommendation model based on MLP.

\begin{figure*}[!ht]
\centering
\includegraphics[scale = 0.52]{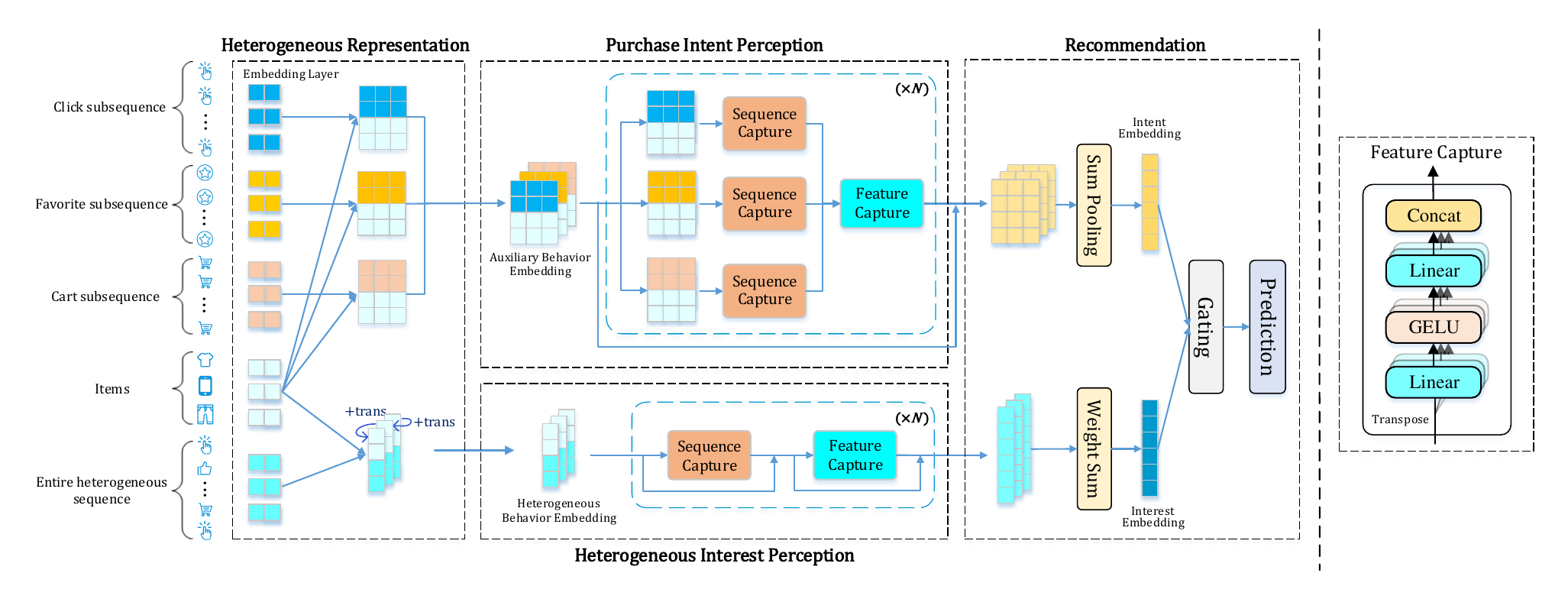}
\caption{ The structure of the Behavior-aware MLP (BMLP) is divided into Heterogeneous Interest Perception (HIP), Purchase Intent Perception (PIP), and recommendation module. Details of the behavior replacement block and the Feature Capture Block (FCB) are shown in the right panel. The Sequence Capture Block (SCB) is a special case of FCB.}
\label{fig:model_fcd}
\end{figure*}


\section{METHODOLOGY}

\label{sec:METHODOLOGY}
\indent In this section,  we {first} formally describe the studied problem and then introduce our BMLP in detail. 

\subsection{Problem Definition}
\indent In heterogeneous sequential recommendation (HSR), our task is to predict the next item likely to be purchased for each user based on a user's historical interaction sequence with different types of behaviors, such as clicks and purchases. {In contrast to multi-task recommendation~\cite{MTDRS}, which aims for balanced performance improvement across all behaviors, our goal is to predict the purchase behavior.}
Without loss of generality, we have some different types of behaviors $b \in \mathcal{B}$ given by a set of users $\mathcal{U}$ to a set of items $\mathcal{I}$. 
In HSR, we define a historical heterogeneous sequence of a user $u$ as $\mathcal{S}_u =\{(i_u^{1}, b_u^{1}),(i_u^2, b_u^2),\ldots,(i_u^{|\mathcal{S}_u|}, b_u^{|\mathcal{S}_u|} )\}$, $i_u^{\cdot} \in \mathcal{I}$; $b_u^{\cdot} \in \mathcal{B}$. 
Our goal is to predict the next purchased item $i_u^{|\mathcal{S}_u| +1}$ of a user $u$ from $\mathcal{U}$ as accurately as possible.

\subsection{Overview of Behavior-Aware MLP}
\indent The overall structure of BMLP is shown in Fig~\ref{fig:model_fcd} and consists of three modules as follows: 
1) The HIP module encodes the entire heterogeneous sequence to ensure the integrity of the sequential information and {capture} fine-grained dependencies. 
It mainly learns the dependencies between {the} items and the representation information {of items and behavior types} separately by transposing the input tensor and using MLP twice. 
2) The PIP module also {firstly} transposes each auxiliary behavior subsequence, then uses {an} MLP to learn the dependencies between the items of each subsequence separately, and finally uses an MLP to fuse all the auxiliary behavior {embeddings}. 
3) The recommendation module adaptively fuses the learned {interest and intent from} two complementary modules. 
\\
\indent As shown in Fig~\ref{fig:model_fcd}, we first encode the item {embeddings} and behavior embeddings of the {entire} heterogeneous sequence and each auxiliary behavior subsequence. 
For the entire heterogeneous sequence, we {further} introduce behavior transition embedding~\cite{Transrec++}. 
{Then, we use the HIP module and PIP module to compute the user's overall heterogeneous interest and purchase intent, respectively.}
{Finally, we fuse them using a gating {component} to obtain the final interest. 
We then use this final interest to calculate the dot product with the item embeddings, resulting in the corresponding item rating predictions for {each} user.}

\subsection{Heterogeneous Interest Perception}
\indent {By modeling the entire {heterogeneous} sequence, it is possible to perceive the overall preferences of the user. 
However, this may suffer from insufficient utilization of different {behavior} information. 
To address this, we introduce the behavior type information and the behavior transition relations in the user sequences.}
\\
First of all, we select the last $L$ (item, behavior) pairs for each user, which is formalized as $\mathcal{S}_u =\{(i_u^{1}, b_u^{1}),(i_u^2, b_u^2),\ldots,(i_u^{L}, b_u^{L} )\}$. 
Notice that $L$ is usually chosen to be a relatively large value, e.g., $L=50$, which is determined according to the size and distribution of a dataset. 
Moreover, we will pad items or behaviors to the beginning of a sequence if the length is shorter than $L$.  
\\
\indent For each item $i_u^t$ and the corresponding behavior $b_u^t$ of a user $u$ in an interaction sequence $\mathcal{S}_u$, since we deal with a sequence with different types of behaviors, in order to distinguish the behavior types of the items that a user interacts with, we use $B_b\in \mathbb{R}^{1\times d} $ to represent the {embedding} of behavior type $b$. 
We assume that the behavior types of two adjacent items in a sequence are dependent. 
{Following~\cite{Transrec++}, we use $\mbox{trans}(b_u^t, b_u^{t+1})\in \mathbb{R}^{1\times d }$ to denote the transition relationship between two consecutive behaviors.}
Then, we combine the behavior embedding and the behavior transition embedding:
\begin{equation}
	M_u^t = B_{b_u^t} + \mbox{trans}(b_u^t, b_u^{t+1} )
\end{equation}
{Next}, we use $V_{i_u^t}$ to represent the {embedding of item} $i_u^t$. 
Finally, we concatenate the item embedding and the behavior {embedding} to obtain a heterogeneous embedding $\mathbf{x}_u^t \in\mathbb{R}^{1\times 2d }$:
\begin{equation}
	\mathbf{x}_u^t = \mbox{concat}(M_u^t, V_{i_u^t})
\end{equation}
{After the above process}, we obtain an input matrix $\mathbf{X}^{(0)}_u=[\mathbf{x}_u^1;\ldots;\mathbf{x}_u^L] \in \mathbb{R}^{L\times 2d}$, where $L$ {is} the length of the interaction  sequence. 
Notice that we now omit the subscript $u$ in $\mathbf{X}_u^{(0)} $ for brevity. 
Then, we feed the input matrix $\mathbf{X}^{(0)}$ into a series of stacked blocks. 
First of all, we feed it into a sequence capture block (SCB) used to capture the sequential information between items, where the output of the $n$-th SCB is as follows:
\begin{equation}
	\mathbf X^{(n-1)}_{\tiny \mbox{SCB}} =\mathbf X^{(n-1)} + \mbox{SCB}\left(\mbox{LayerNorm}(\mathbf X^{(n-1)})^T \right)^T
\end{equation}
\begin{equation}
	\mbox{SCB} (\mathbf{X}^{T}) = \mbox{GELU}( \mathbf{X}^{T} W_{\tiny \mbox{SCB}}^1) W_{\tiny \mbox{SCB}}^2
\end{equation}
where $W_{\tiny \mbox{SCB}}^1 \in \mathbb{R}^{L\times d_t}$ and $W_{\tiny \mbox{SCB}}^2 \in \mathbb{R}^{d_t\times L}$ are trainable matrices, and $d_t$ represents the hidden layer dimension in SCB. 
\\
 \indent {After updating the columns of the input matrix $\mathbf{X}^{(0)}$, we capture the sequential information between the items from another perspective. 
 {Notice that each} column of $\mathbf{X}^{(0)}$ corresponds to $L$ items in a channel.
Therefore, it can perceive the positional relationship between items.}

We use $\mathbf X^{(n - 1)}_{\tiny \mbox{SCB}}$ as input to the feature capture block (FCB), where $\mathbf X^{(n - 1)}_{\tiny \mbox{SCB}}$ is divided into the {$H$ parts} of the hidden layer dimension, i.e., $\mathbf{X}_{ (h-1) \times \frac{2d}{H} +1 : h \times \frac{2d}{H} }$, $h = 1, 2, \ldots, H$.
{It} guides the model to focus on the information in different spaces. 
We feed $\mathbf{X}_{ (h-1) \times \frac{2d}{H} +1: h \times \frac{2d}{H} }  \in \mathbb{R}^{ L \times \frac{2d}{H}}$  into a smaller feature capture block, defined as a head. 
{Finally, we merge and fuse the information of {those $H$ heads}.
Among them,} the $n$-th block is calculated as follows:
\begin{equation}
	\mathbf X^{(n)} = \mathbf X^{(n -1)}_{\tiny \mbox {SCB}} +  \mbox{FCB}(\mbox{LayerNorm}(\mathbf X^{(n -1)}_{\tiny \mbox{SCB}}))
\end{equation}
\begin{equation}
	\mbox{FCB}(\mathbf X) = \mbox{concat}(\ldots, \mbox{head}_h, \ldots)W_{\tiny \mbox{FCB}}^O
\end{equation}
\begin{equation}
	\mbox{head}_h =  \sigma( \mathbf{X}_{ (h-1) \times \frac{2d}{H} +1 : h \times \frac{2d}{H} } W_{\tiny \mbox{FCB}}^1 ) W_{\tiny \mbox{FCB}}^2, h = 1, 2, \ldots, H
\end{equation}
{where $W_{\tiny \mbox{FCB}}^O \in \mathbb{R}^{2d\times 2d}$, $W_{\tiny \mbox{FCB}}^1 \in \mathbb{R}^{\frac{2d}{H}\times d_c}$} and $W_{\tiny \mbox{FCB}}^2 \in \mathbb{R}^{d_c\times \frac{2d}{H}}$ are learnable matrices, $\sigma(\cdot)$ is a sigmoid activation function, and $d_c$ {is} the hidden layer dimension in FCB. 
We can see that SCB is a special case of FCB when $H=1$. 
Notice that SCB and FCB deal with the sequential information and the feature information, respectively.
\indent We take the output matrix $\mathbf X^{(N)} \in \mathbb{R}^{L\times 2d}$. 
Considering the different importance of each item in a sequence, we use a weighting mechanism to aggregate them:
\begin{equation}
	\alpha_t = \frac{\exp(\mathbf{x}_t^{(N)}W_{\alpha})}{\sum_{j=1}^{L}\exp(\mathbf{x}_j^{(N)}W_{\alpha})}
\end{equation}
where $W_{\alpha} \in \mathbb{R}^{2d\times 1}$ is a learnable vector, and  $\alpha_t \in \mathbb{R}$ represents the weight {for the $t$-th item} in the sequence.
Then, the corresponding item {embeddings} are fused by the weights:
\begin{equation}
	\mathbf e_g = \mbox{dropout}(\sum_{t=1}^{L}\alpha_t \mathbf x_t^{(N)})
\end{equation}
\indent In this module, we obtain the corresponding heterogeneous behavior representations by fusing item representations, behavior representations, and behavior transition information. 
{Then,} we utilize the SCB and FCB modules to capture the fine-grained long-term interests of users.
\subsection{Purchase Intent Perception}

{In the previous module, we fuse the behavior embedding and the item embedding to obtain the global heterogeneous interest of each user}. 
However, there are some limitations in this approach. 
Firstly, the HIP module cannot distinguish the behavior type of the next predicted item. 
Although the long-term heterogeneous interest can help us understand a user's general preferences, it does not take into account the specificity of his or her purchase behavior. 
This is not consistent with our goal of predicting the next item that a user may purchase {in the future}. 
Secondly, in an e-commerce scenario, a user tends to click some similar items or mark them as favorite before purchasing. 
This means that a recent auxiliary behavior has a greater influence on a user's {purchase behavior}.
To address this, we further design a purchase intent perception module. 
This module takes into account the recent auxiliary {behaviors} to better understand a user's current purchase intent.
\\
\indent 
The recent auxiliary behaviors (e.g., clicks, favorites) of {a user} reflect the short-term purchase intent of {the user}, so we choose the most recent $L'$ {items w.r.t. each} auxiliary behavior as a subsequence, where $L'$ is relatively small (e.g., $L' = 5$). 
Earlier auxiliary behaviors usually have little effect {on the} current purchase intent because the purchase has already occurred or {the user's} interest has shifted.
\\
\indent  For each auxiliary behavior, we concatenate the embedding of the item and {the behavior type} as the input of the purchase intent perception module:
\begin{equation}
	\mathbf h_{b_u}^t = \mbox{concat}(	 B_{b_u^t} ,V_{i_u^t})
\end{equation}
where  $ B_{b_u^t}\in \mathbb{R}^{1\times d} $ represents the {embedding} of the auxiliary behavior $b_u^t$ and $V_{i_u^t}\in  \mathbb{R}^{1\times d} $ represents the {embedding} of the item $i_u^t$. 
So, we obtain an input tensor $\mathcal{H}^{(0)}_{u}=[\mathbf H_{b_u}^1;\ldots;\mathbf H_{b_u}^{m}] \in \mathbb{R}^{{L'}\times m \times 2d}$. 
Notice that $m$ indicates the number of auxiliary behaviors. 
Similarly, we omit the subscript $u$ in $\mathcal{H}_{u}^{(0)} $ for brevity. 
\\
\indent Firstly, we transpose the embedding matrix for each auxiliary behavior. 
Then we feed them into SCB and FCB, where the output is as follows:
\begin{equation}
	\mathcal {H}^{(n -1)}_{s} =\mbox{stack} \left(\mbox{SCB}\left(\mathbf H_{b_u^1}^{{(n-1)}^T}\right)^T, ... ,\mbox{SCB}\left(\mathbf H_{b_u^m}^{{(n-1)}^T}\right)^T\right)
\end{equation}
\begin{equation}
	\mathcal {H}^{(n)} = \mathcal {H}^{(n -1)}  + \mbox{FCB}(\mbox{LayerNorm}(\mathcal {H}^{(n -1)}_{s}))
\end{equation}
where the $\mbox{SCB}(\cdot)$ and $\mbox{FCB}(\cdot)$ {are} the same as that in the heterogeneous interest learning module. 
The difference between the HIP module and the PIP module is the heterogeneity and length of the processed sequence.
We aggregate the last item {embedding of} each auxiliary behavior as the short-term purchase intent:
\begin{equation}
	\mathbf e_l = \mbox{mean}(\mathcal {H}_{L'}^{(N)})
\end{equation}
where $\mathbf e_l \in \mathbb{R}^{1\times 2d}$, and $\mathcal {H}_{L'}^{(N)}\in \mathbb{R}^{m\times 2d}$ denotes the last item representation of {the}  auxiliary behaviors.
\subsection{Gating}
\indent {We combine the local purchase intent $\mathbf e_l$ and the global heterogeneous interest $\mathbf e_g$ as follows:}
\begin{equation}
	\mathbf g =  \sigma(\mathbf e_g  W_g +  \mathbf e_{l} W_l+ \mathbf{b}_g)
\end{equation}
\begin{equation}
	\mathbf z = \mathbf g \otimes \mathbf e_g + (1-\mathbf g) \otimes \mathbf e_{l}
\end{equation}
where $\otimes$ is the element-wise product, $W_g \in \mathbb{R}^{2d\times 2d}$, $W_l \in \mathbb{R}^{2d\times 2d}$ and $\mathbf{b}_g \in \mathbb{R}^{1\times 2d}$ are learnable weights and biases, and $\sigma(\cdot)$ is a sigmoid activation function to constrain the value of each entry in $\mathbf g \in \mathbb{R}^{1\times 2d}$ to $(0,1)$.
Finally, we use a fully connected layer to make the prediction:
\begin{equation}
	\hat r_{t+1} = \mbox{softmax}(\mathbf z W_r + \mathbf{b}_r)
\end{equation}
where $W_r \in \mathbb{R}^{2d\times |\mathcal{I}|}$ is a learnable matrix, $\mathbf{b}_r \in \mathbb{R}^{1\times |\mathcal{I}|}$ is the bias, and $\hat r_{t+1} \in \mathbb{R}^{1\times |\mathcal{I}|}$ {contains} the predicted score on each item.
\subsection{Loss Function}
\indent {The loss function of the model is as follows}:

\begin{equation}
	\mathcal{L} = -\sum_{s\in \mathcal{S}} \sum_{j \in \mathcal{I}} y_{sj} \log(\hat r_{t+1,j}) + (1-y_{sj}) \log (1-\hat r_{t+1,j})
\end{equation}
where $y_{sj}=1$ only if an item $j$ is a {truly interacted item in} the sequence $s$ at timestamp $t+1$, and $y_{sj}=0$ otherwise.

\subsection{Discussions}
\label{sec:discussions}
\indent In this subsection, we discuss three existing MLP-based sequential recommendation methods.
\begin{itemize}[leftmargin=*]
	\item MOI-Mixer~\cite{MOI-Mixer}: It is the first work to apply MLP-Mixer~\cite{MLP-Mixer} in a recommender system, which introduces higher-order interactions in the MLP layer. {Notice that} we introduce the multi-head mechanism in the MLP layer to learn the feature information in different subspaces, and capture behavior information with {different} {granularities} at the same time.
	\item FMLP-Rec~\cite{2022FMLP}:  It is an improved method based on MLP-Mixer~\cite{MLP-Mixer}. The difference is that it passes a filter layer after encoding the items, which reduces the noise. Our BMLP extracts {user interest and intent from} heterogeneous sequences, which can also reduce noise to a certain extent because of their complementarity.
	\item MLP4Rec~\cite{MLP4Rec}: It is also an improved method based on MLP-Mixer~\cite{MLP-Mixer}, which introduces attributes such as {categories and brands} of items and expands them into a new dimension. Specifically, the input of MLP-Mixer is a 2-D matrix, {while} the input of MLP4Rec is a 3-D tensor.
\end{itemize}
\indent {{We can see} that the MLP architecture has rarely been exploited in recommendation systems}. 
Notice that the structure of MLP is insensitive to the sequential information. 
{Our BMLP captures the sequential pattern} and achieves better performance in sequential recommendation {mainly with the following two reasons}:
1) the sliding window training approach, which allows the model to capture the sequential information in the training process; 
{and 2)} the transposition of the input matrix, which {allows} the MLP to capture the dependencies between items.

\section{Experiments}

\label{sec:experiments}
\indent In this section, we study the effectiveness of our proposed model by conducting extensive experiments on four datasets. 
We first introduce the experimental setup in detail, including data processing, baselines, and evaluation metrics, and then focus on answering the following {seven} research questions.
\begin{itemize}[leftmargin=*]
	\item $\mathbf{RQ1:}$ How does our proposed model perform compared with the state-of-the-art models?
	\item $\mathbf{RQ2:}$ Why is our BMLP simpler and more efficient compared with the existing methods?
	\item $\mathbf{RQ3:}$ How will behavior diversity affect the performance of the model?
	\item $\mathbf{RQ4:}$ How do the components, i.e., FCB, SCB, and PIP,  affect the performance of the model?
	\item $\mathbf{RQ5:}$ How does the PIP module perceives a user's purchase intent?
    {\item $\mathbf{RQ6:}$ How do the heterogeneous non-sequential recommendation methods perform?}
    
	\item $\mathbf{RQ7:}$ What is the impact of the values of the hyperparameters on the model?

\end{itemize}

\subsection{Experimental Setting}
\label{sec:experiment setting}

\subsubsection{Datasets}
We conduct experiments on four datasets, i.e., RecSys Challenge 2015 (Rec15)\footnote{\url{https://recsys.acm.org/recsys15/challenge/}}, Tmall at IJCAI-15 Contest (Tmall)\footnote{\url{https://tianchi.aliyun.com/dataset/dataDetail?dataId=42}}, MovieLens 1M (ML1M)\footnote{\url{https://grouplens.org/datasets/movielens/1m/}}, and Taobao user behaviors (UB)\footnote{\url{https://tianchi.aliyun.com/dataset/dataDetail?dataId=649}}. 
We process these datasets as follows: 
1) For duplicated (user, item, behavior) records, we only keep the record with the earliest time. 
2) For each user's interaction sequence, we use the last two purchases for validation and {testing}, respectively, and retain the auxiliary behaviors between these two purchases for final performance evaluation. 
3) We remove the cold-start items in the validation and test sets. 
\\
Then we {conduct} some specific processing for each dataset.

\vspace{0.05in}

\noindent
\textbf{Rec15}. Rec15 was released by the RecSys 2015 competition, which contains 9,249,729 users, 52,739 items, and 34,154,697 interaction records. 
We preprocess it as follows:
1) Delete the items that have been purchased fewer than $5$ times. 
2) Delete the users who have purchased items fewer than $5$ times.

\noindent
\textbf{Tmall}. The Tmall dataset comes from the Tmall app and contains records of user interactions before and on the day of “Double 11”. 
The original data contains 424,170 users, 1,090,390 items, four types of behaviors, and 54,925,330 interaction records. 
We preprocess it as follows:
1) Delete all the records on the day of ``Double 11" to avoid impulsive consumption and the disruption caused by the promotional activities on that day. 
2) Delete the items that have been purchased fewer than $20$ times. 
3) Delete the users who have purchased items fewer than $10$ times.

\noindent
\textbf{ML1M}. A dataset of movie ratings with 1 million ratings assigned by 6,000 users to 4,000 movies. 
We preprocess it as follows:
1) A rating of 5 is simulated as a purchase, and a rating smaller than 5 is taken as an auxiliary behavior~\cite{Transrec++}. 
2) Delete the items that have been purchased fewer than $5$ times. 
3) Delete the users who have purchased items fewer than $5$ times. 

\noindent
\textbf{UB}. 
A Taobao behavior dataset provided by Alibaba, which contains 987,994 users, 4,162,024 items, and 100,150,807 interaction records. 
We preprocess it as follows:
1) Delete the items that have been purchased fewer than $10$ times. 
2) Delete the users who have purchased items fewer than $5$ times.

\indent The statistics of the processed datasets are summarized in Table~\ref{tb:statistic}. {In addition, we observe that some items have been examined in the input sequence while others have not. Considering the impact of the auxiliary behaviors on the target behavior and the fact that users in real-world recommendation scenarios are likely to purchase items they have previously clicked, the model recommends both examined and unexamined items. Subsequently, we conduct an in-depth analysis of this aspect in Section 4.7.}
{The processed datasets, source code of our BMLP and the scripts used in the experiments are publicly available at https://csse.szu.edu.cn/staff/panwk/publications/BMLP/.}

\doublerulesep 0.1pt
\begin{table}[h]
\begin{footnotesize}
\caption{Statistics of the four processed datasets used in the experiments.} \label{tb:statistic}
\resizebox{\linewidth}{!}{
\begin{tabular}{p{0.8cm}p{0.6cm}p{0.8cm}p{1.2cm}p{3.5cm}}
\hline\hline\noalign{\smallskip}
    Datasets & \#Users & \#Items & \#Interactions & \#Behaviors \\
\noalign{\smallskip} \hline
					\multirow{1}{*}{Rec15} &  36,917 & 9,620 & 679,704 & \{Click, Buy\} \\
					
					\multirow{1}{*}{Tmall} & 17,209 & 16,162 & 1,251,829 & \{Click, Favorite, Buy\}\\
					
					\multirow{1}{*}{ML1M} & 5,645 & 2,357 & 885,598 & \{Click, Buy\} (simulated) \\
					
					\multirow{1}{*}{UB} & 20,858 & 30,718 & 782,297&  \{Click, Cart, Favorite, Buy\}\\

					\hline\hline

\end{tabular}
}
\end{footnotesize}
\end{table}

	
		
					
					
					
					
					

					

\subsubsection{Evaluation Metrics}
We adopt two widely used metrics to compare our proposed model with the baselines,  i.e., hit ratio {(HR@$k$)} and normalized discounted cumulative gain (NDCG@$k$), where $k$ is chosen from $\{10, 20\}$. 
HR considers whether the test items appear in a recommendation list, while NDCG is more concerned with the positions of the test items.
For a heterogeneous sequence of a {particular} user, we take the last purchased item as the test data. 
{According to} the score of each item calculated in Eq.(16), we select $k$ items with the highest scores as a recommendation list.

\subsubsection{Baselines}
We include some competitive baseline methods for both single-behavior sequential recommendation (i.e., Caser, GRU4Rec, SASRec, SRGNN and BERT4Rec) and heterogeneous sequential recommendation (i.e., RLBL, RIB, BINN, and MSR).
\begin{itemize}[leftmargin=*]
	\item Caser~\cite{Caser}: It constructs users' behavior embeddings into matrices and uses CNN to capture multi-order relationships between items.
	\item GRU4Rec~\cite{GRU4Rec}: It takes a user's behaviors as a sequence and uses GRU to learn the dependencies between items.
	\item SASRec~\cite{SASRec}: It is an attention-based sequential recommendation model that captures the importance of items in the sequence.
	\item SRGNN~\cite{SRGNN}: It captures complex transition relationships between {the items by constructing a graph from a sequence}.
	\item BERT4Rec~\cite{bert4rec}: It predicts the items {masked in the sequence by utilizing the context around them}.
\end{itemize}
\begin{itemize}[leftmargin=*]
	\item RLBL~\cite{RLBL}: It is a circular log bilinear model that utilizes transition matrices to model the information of behavior types in a sequence.
	\item RIB~\cite{RIB}: It adds behavior embedding into the input layer and uses GRU and attention layers to capture the micro-behaviors in a sequence.
	\item BINN~\cite{BINN}: It uses a dual LSTM model to learn a user's long-term static interest and short-term dynamic interests and adds behavior information to LSTM.
	\item MSR~\cite{MSR}: It uses graph neural networks to capture relationships between items and uses GRU to model {heterogeneous} sequences at the same time.
\end{itemize}

\indent Moreover, since our BMLP is based on MLP, we {also} include three closely related state-of-the-art MLP-based sequential recommendation methods (i.e., MLP-Mixer, MOI-Mixer, and FMLP-Rec).
\begin{itemize}[leftmargin=*]
	\item MLP-Mixer\footnote{\url{https://github.com/lucidrains/mlp-mixer-pytorch}}~\cite{MLP-Mixer}: It is a new framework {in computer vision} that uses two independent MLPs to deal with classification tasks, which performs {comparably} to CNN and Transformer.
	\item MOI-Mixer~\cite{MOI-Mixer}: It is improved based on MLP-Mixer~\cite{MLP-Mixer} and proposes a multi-order sequential recommendation model.
	\item FMLP-Rec\footnote{\url{https://github.com/Woeee/FMLP-Rec}}~\cite{2022FMLP}:  It introduces a filter to reduce {the noise} after encoding the items.
\end{itemize}

\subsubsection{Experimental details}
For our proposed model BMLP and all the baselines, we use grid search to select the best values of the parameters on the validation data.
{
Specifically, we set the initial learning rate as $0.01$ and the batch sample size as $512$.
The hidden layer dimension is selected from $\{64, 128, 256\}$.
We fix the sequence length $L=50$ and the auxiliary behavior subsequence length $L'=5$.
The number of heads and blocks are selected from $\{1, 2, 4, 8\}$ and $\{1, 2, 3\}$, respectively.
To avoid overfitting, we choose the dropout rate from $\{0.2, 0.3, 0.4, 0.5\}$ and the regularization parameters from $\{0.0001, 0.001, 0.01, 0.1\}$. 
}
For Caser, we set the number of vertical and horizontal filters to $4$ and $16$, respectively, and choose the height of the horizontal filters from $\{2, 3, 4\}$. 
For RLBL, we select the window size from $\{1, 2, 3, 4, 5\}$. 
We use Adam for optimization.
{ Finally, we evaluate the performance of our BMLP and all the baselines on the test data.}

\begin{table*}[!htbp]
    \caption{The overall performance of our BMLP and twelve baselines. Notice that the best results are marked in bold and the second best are underlined.}
    \label{tb:main-results}
	\renewcommand\arraystretch{1.5}
	\begin{center}
		\setlength{\tabcolsep}{1.5mm}{
			\scriptsize
			\begin{tabular}{l | l  c c c c c| c c c c|c c c |c}\hline\hline
				
				Dataset & Metric & Caser & GRU4Rec & SRGNN & BERT4Rec & SASRec & RLBL & RIB  & BINN & MSR & MLP-Mixer & MOI-Mixer & FMLP & BMLP\\ \hline
    
				
				\multirow{4}{*}{Rec15}
				& HR@10
				& 0.461
				& 0.437
				& 0.483
				& 0.389
				& 0.437
				& 0.331
				& 0.387
				& 0.468
				& \underline{0.605}
			
				& 0.478
				& 0.482
				& {0.541}
				& \bf 0.618 \\
				
				& NDCG@10
				& 0.247
				& 0.249
				& {0.285}
				& 0.196
				& 0.248
				& 0.160
				& 0.213
				& 0.257
				& \underline{0.378}
				
				& 0.259
				& 0.260
				& 0.255
				& \bf 0.407 \\
				
				& HR@20
				& 0.612
				& 0.561
				& 0.608
				& 0.535
				& 0.566
				& 0.459
				& 0.521
				& 0.611
				& \underline{0.715}
				
				& 0.635
				& 0.640
				& 0.672
				& \bf 0.726 \\
				
				& NDCG@20
				& 0.287
				& 0.281
				& 0.317
				& 0.233
				& 0.283
				& 0.191
				& 0.248
				& 0.293
				& \underline{0.408}
				
				& 0.298
				& 0.299
				& 0.286
				& \bf 0.434\\
				\hline
				
				\multirow{4}{*}{Tmall}
				& HR@10
				& 0.098
				& 0.352
				& 0.274
				& 0.168
				& 0.369
				& 0.243
				& 0.269
				& 0.298
				& 0.298
			
				& 0.292
				& 0.317
				& \underline{0.381}
				& \bf 0.426 \\
				
				& NDCG@10
				& 0.076
				& 0.239
				& 0.187
				& 0.096
				& 0.281
				& 0.163
				& 0.188
				& 0.199
				& 0.195
			
				& 0.203
				& 0.215
				& \underline{0.309}
				& \bf{0.315} \\
				
				& HR@20
				& 0.122
				& 0.411
				& 0.319
				& 0.225
				& 0.415
				& 0.281
				& 0.306
				& 0.349
				& 0.348
				
				& 0.361
				& 0.369
				& \underline{0.421}
				& \bf 0.472 \\
				
				& NDCG@20
				& 0.081
				& 0.252
				& 0.196
				& 0.109
				& 0.292
				& 0.181
				& 0.202
				& 0.213
				& 0.211
			
				& 0.221
				& 0.228
				& \underline{0.315}
				& \bf{0.326} \\
				\hline
				
				\multirow{4}{*}{ML1M}
				
				& HR@10
				& 0.215
				& 0.268
				& 0.249
				& 0.145
				& 0.243
				& 0.205
				& 0.229
				& \underline{0.273}
				& 0.265
			
				& 0.264
				& 0.269
				& 0.116
				&  \bf 0.304 \\
				
				& NDCG@10
				& 0.119
				& 0.152
				& 0.141
				& 0.072
				& 0.132
				& 0.102
				& 0.122
				& \underline{0.155}
				& 0.150
			
				& 0.145
				& 0.151
				& 0.058
				& \bf 0.179 \\
				
				& HR@20
				& 0.311
				& 0.361
				& 0.343
				& 0.237
				& 0.345
				& 0.302
				& 0.335
				& 0.376
				& 0.373
				
				& 0.367
				& \underline{0.378}
				& 0.169
				& \bf 0.407 \\
				
				& NDCG@20
				& 0.142
				& 0.180
				& 0.167
				& 0.095
				& 0.159
				& 0.129
				& 0.150
				& \underline{0.181}
				& 0.176
				
				& 0.171
				& 0.179
				& 0.068
				& \bf 0.204 \\
				
				\hline
				
				\multirow{4}{*}{UB}
				& HR@10
				& 0.126
				& 0.171
				& 0.142
				& 0.098
				& \underline{0.181}
				& 0.093
				& 0.132
				& 0.141
				& 0.172
			
				& 0.145
				& 0.144
				& 0.158
				& \bf 0.200 \\
				
				& NDCG@10
				& 0.086
				& 0.104
				& 0.083
				& 0.055
				& \underline{0.106}
				& 0.055
				& 0.080
				& 0.083
				& 0.099
			
				& 0.085
				& 0.085
				& 0.103
				& \bf 0.114 \\
				
				& HR@20
				& 0.152
				& 0.219
				& 0.186
				& 0.138
				& \underline{0.238}
				& 0.122
				& 0.168
				& 0.189
				& 0.225
			
				& 0.187
				& 0.187
				& 0.192
				& \bf 0.254 \\
				
				& NDCG@20
				& 0.085
				& 0.116
				& 0.092
				& 0.063
				& \underline{0.121}
				& 0.062
				& 0.089
				& 0.097
				& 0.117
				
				& 0.099
				& 0.099
				& 0.113
				& \bf 0.130\\
				\hline\hline
				
		\end{tabular}
  }

	\end{center}
\end{table*}

\subsection{Overall Performance (RQ1)}
In this section, we compare our model with the baselines on four datasets {and show the results} in Table~\ref{tb:main-results}. 
According to the experimental results, we {can} obtain the following observations.
\begin{itemize}[leftmargin=*]
\item Our BMLP achieves a significant improvement compared with the baselines on all {the four} datasets, which clearly shows the effectiveness of our model. 
The MLP-based {models} can beat SASRec and GRU4Rec on RC15, Tmall, and ML1M. 
It demonstrates that a pure MLP model has {the} potential {for} sequential recommendation. 
Compared with MLP-based sequential recommendation methods, our BMLP shows a significant improvement in recommendation performance, which indicates that our model is able to capture the multi-behavioral dependencies more comprehensively. 
\item MSR and SRGNN are GNN-based models. MSR performs better than SRGNN on all the datasets. 
This shows that multi-behavior information contributes to {the performance} improvement. 
In contrast, BINN performs worse than GRU4Rec on some datasets, which means that BINN may not capture the dependencies between behaviors well.
\item Caser is more prominent on Rec15, for which we speculate that users' short-term interests may dominate on Rec15. 
FMLP-Rec performs poorly on ML1M but well on the other datasets. 
{It shows that although a dense data helps capture users' interests, it is also necessary to model them in a fine-grained manner.} Otherwise, it may lose {some} important information using the filter layers.
{\item The graph-based methods SRGNN and MSR perform well on the Rec15 dataset, but not as well on the other datasets. The main reason is that the graph-based approach is more effective for datasets with short sequences. As shown in Table~\ref{tb:statistic}, the Rec15 dataset has the shortest user sequences.}
\end{itemize}
\subsection{Efficiency Analysis (RQ2)}
\label{sec:Time Complexity Analysis}
\indent In this subsection, we compare our model with two classical sequential recommendation models, i.e., GRU4Rec and SASRec, in terms of time complexity. 
For GRU4Rec, the time complexity of computing an item is $O(6d^2)$. 
Hence, for $L$ items, the overall time complexity is $O(6Ld^2)$. 
SASRec has the following three operations: 
1) mapping the hidden representations three times to obtain the query, key, and values; 
2) calculating the attention scores, performing weighted sum; 
3) passing through the fully connected layer. 
For the fully connected layer, the hidden unit is a constant independent of the input and output units, and the number of input unit is equal to the output unit. 
Therefore, their time complexity are $O(3Ld^2)$, $O(dL^2 + Ld^2)$ and $O(2Ld)$, where $L$ is the length of the sequence and $d$ is the dimension of the representation. 
We can obtain the overall time complexity of SASRec as $O(4Ld^2 + dL^2 + 2Ld)$. 
For {our BMLP, {the} main} modules are PIP and HIP, and their operations are fully connected layers. 
Therefore, their time {complexities} are $O(4dL)$ and $O(4mdL')$, where $m$ denotes the number of auxiliary behaviors and $L'$ denotes the length of the auxiliary behavior subsequence. 
We set the number of {heads} to 1 for brevity.  
Based on the above analysis, we can conclude that {our BMLP is of} a lower time complexity.


Experiments are conducted on Tmall with one Tesla V100 GPU and Intel(R) Xeon(R) Gold 6230R CPU @2.10GHz machine for all {the} compared models. 
Results of the {the} training time and performance metrics for {our} BMLP and baselines are shown in Table 3 and Table 4, respectively. 
We have the following observations:

\doublerulesep 0.06pt
\begin{table}[h]
\begin{footnotesize}
\caption{Results of training efficiency on Tmall.} \label{tab:efficien1}
\begin{tabular}{p{1.3cm}p{1.6cm}p{1.6cm}p{1.8cm}}
\hline\hline\noalign{\smallskip}
Model & Time/Epoch & \#Epoch & \#Training Time \\ 

\noalign{\smallskip} 		\hline	
	\multirow{1}{*}{SASRec} & 22 & 10m & 3h40m \\				
	\multirow{1}{*}{GRU4Rec} & 13 & 12m & 2h36h\\
	
	\multirow{1}{*}{Caser} & 15 & 8m & 2h\\
	
	\multirow{1}{*}{SRGNN} & 11 & 17m & 3h7m\\
	
	\multirow{1}{*}{BINN} & 12 & 22m & 4h24m\\
	\hline	
	\multirow{1}{*}{BMLP} & \textbf{30} & \textbf{3m} & \textbf{1h30m} \\
\hline\hline
\end{tabular}
\end{footnotesize}
\end{table}

\doublerulesep 0.06pt
\begin{table}[!htbp]
\begin{footnotesize}
\caption{Efficiency and performance comparison on Tmall.}
\label{tab:efficien2}
\begin{tabular}{p{1.3cm}p{1.8cm}p{1.6cm}p{1.6cm}}
\hline\hline\noalign{\smallskip}
Model & \#Training Time & \#HR@10 & \#NDCG@10 \\ 

\noalign{\smallskip} 		\hline	
	\multirow{1}{*}{SASRec} & 3h40m & 0.362 & 0.277 \\				
	\multirow{1}{*}{GRU4Rec} & 2h36h & 0.349 & 0.237\\
	\multirow{1}{*}{Caser} & 2h & 0.094 & 0.073\\
	\multirow{1}{*}{SRGNN} & 3h7m & 0.268 & 0.183\\
	\multirow{1}{*}{BINN} & 4h24m & 0.293 & 0.196\\
	\hline				
	\multirow{1}{*}{BMLP} & \textbf{1h30m} & \textbf{0.426} & \textbf{0.315} \\
\hline\hline
\end{tabular}
\end{footnotesize}
\end{table}

(1)
Table~\ref{tab:efficien1} shows that {our} BMLP takes only three minutes for each epoch.
{It} is much less than other models, leading to only 90 minutes for the whole training process. 
{Eventually}, our BMLP achieves around 3x and 2x speedup compared with BINN and SRGNN, respectively, demonstrating the superiority of our BMLP in training efficiency.

(2)
Table~\ref{tab:efficien2} shows the training time when the models converge {and} the performance of the corresponding models. 
We can see that our BMLP can achieve better performance with less time, which further demonstrates the higher efficiency of {our} BMLP in comparison with the other baseline models.

\subsection{Study of Behavior Types (RQ3)}
\indent In this subsection, we mainly discuss how heterogeneous behaviors affect the recommendation performance of the model. 
We conduct four comparative experiments on each dataset.
\begin{itemize}[leftmargin=*]
	\item \textbf{S}: {We ignore the behavior information, i.e., we do not take into account the specific behavior associated with each item.}
	\item \textbf{B}: {We introduce the behavior type by concatenating it with the item feature. 
	Specifically, in Eq.(1), we only consider the behavior type information (represented as $B_{b_u^t}$) in the concatenation step.}
	\item \textbf{T}: We introduce the behavior transition relationship. 
	Specifically, we encode the behavior transition of two adjacent items, e.g., from purchase to click, and finally concatenate it with the item feature. 
	{In Eq.(1), we only consider the behavior transition relationship (represented as $\mbox{trans}(b_u^t, b_u^{t+1}$).}
	\item \textbf{B+T (i.e., BMLP)}: It is a combination of \textbf{B} and \textbf{T}. 
	Specifically, we add the features of the behavior types and behavior transition relationships as the overall behavior representations.
\end{itemize}

\begin{figure}[!ht]
\centering
\includegraphics[scale = 0.32]{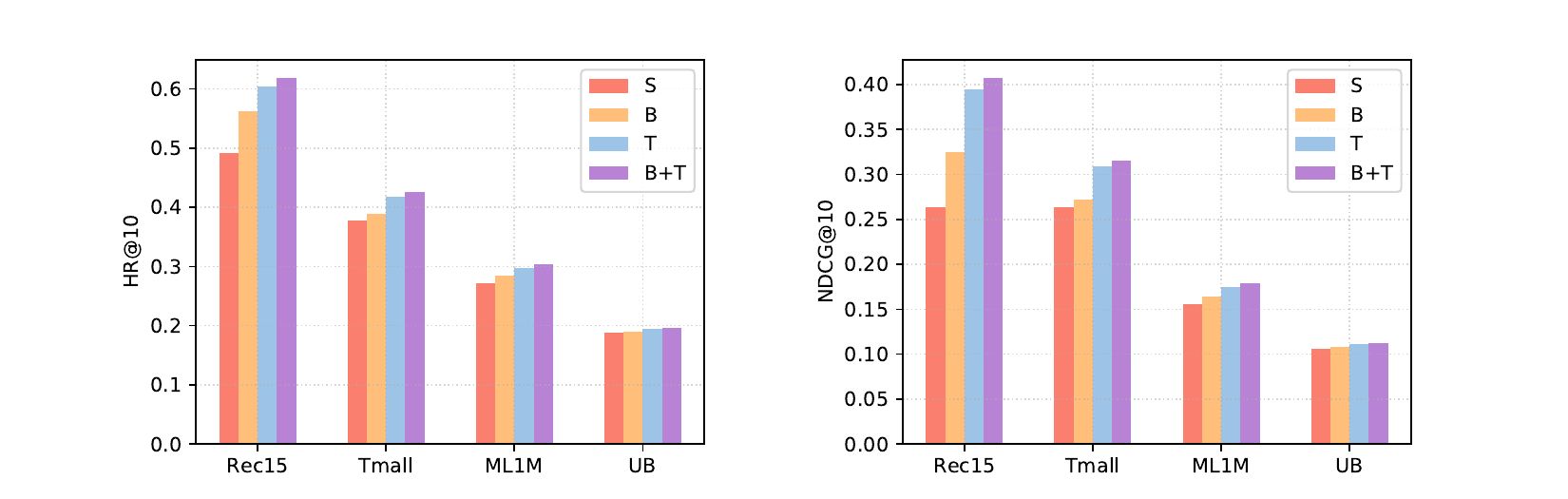}
\caption{ The effectiveness of four different approaches of modeling the behavior types in our BMLP on four datasets. Notice that 'S', 'B', and 'T' denote modeling the sequences without the behavior types, with behavior types, and with behavior transition relationships, respectively.}

\label{fig:behavior}
\end{figure}


\indent From the results in Fig \ref{fig:behavior}, we can have  the following observations: 
1) If the behavior types are ignored, the performance of our model drops significantly. 
2) The introduction of the behavior transition relationship has a greater performance improvement than the introduction of the behavior types. 
Intuitively, in a real e-commerce scenario, a user purchases an item, and later {the user} may click another in the next step. 
Our behavior transition component can capture this micro transition from purchase to click. 
Therefore, the behavior transition relationships may better reflect the diversity and dependencies of the behaviors. 
3) The results of our BMLP in \textbf{B+T} show that the combination of behavior types and behavior {transition} relations further improves {the performance of our model.
Therefore, }we can see that the behavior types and the behavior transition relationships are complementary.

\subsection{Ablation Study (RQ4)}
\indent {To verify the contribution of each component to the overall performance, i.e., the effectiveness of SCB, FCB, and the impact of modeling auxiliary behavior subsequences or heterogeneous sequences, we conduct ablation studies.}
\begin{itemize}[leftmargin=*]
	\item {BMLP $w/o$ FCB}: We remove FCB. The variant of the model is insensitive to the feature information.
	\item{BMLP $w/o$ SCB}: We remove SCB. The variant of this model does not capture the dependencies between items.
	\item {BMLP $w/o$ PIP}: In order to verify the effectiveness of the PIP module, we use the HIP module alone as an ablation study. It does not tap changes {in} local interest.
	\item {BMLP $w/o$ HIP}: We remove HIP to explore the performance impact of just using the {subsequences of auxiliary behaviors.}
\end{itemize}

\doublerulesep 0.1pt
\begin{table}[!h]
	\caption{Results of the ablation studies on four datasets.}
	
	\label{tb:ablation-study}
	\renewcommand\arraystretch{1.5}
	\begin{center}
		\setlength{\tabcolsep}{1.4mm}{
			\scriptsize
			\begin{tabular}{l | c c  |c c | c c| c c }
				
				\hline\hline
				
				&  \multicolumn{2}{c|}{Rec15}  & \multicolumn{2}{c|}{Tmall} &  \multicolumn{2}{c|}{ML1M}	&\multicolumn{2}{c}{UB}\\ 
				
				&HR & NDCG &  HR& NDCG  &HR & NDCG & HR & NDCG \\
				
				\hline	
				\multirow{1}{*}{$w/o$ FCB}
				
				& {0.581}
				& {0.406}
				&0.398
				&0.301
				&0.277
				&0.160
				&0.175
				&0.099\\

				\multirow{1}{*}{$w/o$ SCB}
				
				&{0.490}
				&{0.264}
				&0.328
				&0.221
				& 0.237
				& 0.131
				&0.172
				&0.098 \\

				
				
				\multirow{1}{*}{$w/o$ PIP}
				
				&{0.567}
				&{0.325}
				&0.385
				&0.283
				&\bf 0.308
				&\bf 0.282
				&0.168
				&0.093 \\

				\multirow{1}{*}{$w/o$ HIP}
				
				& {0.401}
				& {0.183}
				&0.274
				&0.177
				&0.194
				&0.104
				&0.145
				&0.081 \\
				\hline
				
				\multirow{1}{*}{BMLP}
				
				& \bf{0.618}
				& \bf{0.407}
				&\bf 0.426
				&\bf 0.315
				& 0.304
				& 0.179
				&\bf 0.200
				&\bf 0.114 \\
				\hline\hline
			\end{tabular}
		}
	\end{center}
\end{table}

The results of the above variants are shown in Table~\ref{tb:ablation-study}. 
We can obtain the following observations: 
1) When any of the components are removed, the performance of the model degrades. 
It shows the effectiveness of each component. 
2) {The performance drops more significantly after removing SCB comparing with that of removing FCB.}
This shows that SCB can capture the sequential patterns better, {i.e.,} the sequential information is relatively more important. 
3) Firstly, the variant performs better when removing the PIP module on ML1M. 
Notice that the auxiliary and purchase behaviors do not exist in ML1M because it is a simulated data. 
So the PIP module is redundant to it. 
Secondly, we observe that the PIP module has a significant performance boost on the other datasets, with an increase of 6\% on Rec15 and about 4\% on Tmall and UB. 
Finally, according to the previous analysis of the results {of} Caser, we can see that short-term interest may {dominate} on Rec15. 
Intuitively, in real e-commerce scenarios, users often have some auxiliary behaviors about an item before purchasing it. 
Therefore, it can perceive this purchase intent sensitively {as an auxiliary module.} 4) The performance degradation is particularly significant when HIP is removed. 
This indicates that capturing the purchase intent through auxiliary behavior subsequences alone is {insufficient. }
This also reflects that the sequential integrity is {essential}. 

To further verify the ability of the SCB module to capture the sequential information, we design a comparison experiment by replacing the SCB module with {a GRU module and a self-attention module.}


\doublerulesep 0.1pt
\begin{table}[!h]
	\caption{Results of comparison experiments on four data sets.}
	
	\label{tb:ablation-study-gru-att}
	\renewcommand\arraystretch{1.5}
	\begin{center}
		\setlength{\tabcolsep}{1.4mm}{
			\scriptsize
			\begin{tabular}{l | c c  |c c | c c| c c }
				
				\hline\hline
				
				&  \multicolumn{2}{c|}{Rec15}  & \multicolumn{2}{c|}{Tmall} &  \multicolumn{2}{c|}{ML1M}	&\multicolumn{2}{c}{UB}\\ 
				
				&HR & NDCG &  HR& NDCG  &HR & NDCG & HR & NDCG \\
				
				\hline	
				\multirow{1}{*}{BMLP\_{GRU}}
				
				& 0.454	& 0.258 & 0.396	& 0.288 & 0.281 & 0.162 & 0.189 &0.107\\
							
				\multirow{1}{*}{BMLP\_{Att}}
				
				& 0.523
				& {0.296}
				& 0.419

				& 0.311
				& 0.293	

				& 0.169
				&\bf 0.209	

				& \bf0.118 \\
		
				\hline
				
				\multirow{1}{*}{BMLP\_{MLP}}
				&\bf0.618	&\bf0.407&\bf0.426&	\bf0.315&\bf0.304 &	\bf0.179 & 0.200 &	0.114\\
				
				\hline\hline
			\end{tabular}
		}
	\end{center}
\end{table}
{The results of the comparison experiments are presented in Table~\ref{tb:ablation-study-gru-att}, and the key observations are summarized below. 
Firstly, after replacing the SCB module with the GRU {or the self-attention module}, our BMLP(MLP) still outperforms BMP(GRU) and BMLP(Att) on three datasets. 
This demonstrates the effectiveness of the SCB module in capturing the sequential information. 
Secondly, by integrating the GRU {or the self-attention module} into our BMLP, it achieves a significant improvement in overall performance compared with GRU4Rec or SASRec. 
This highlights the versatility and generality of our proposed architecture.}

\subsection{Purchase Intent Analysis (RQ5)}
\begin{figure}[!ht]
\centering
\includegraphics[scale = 0.35]{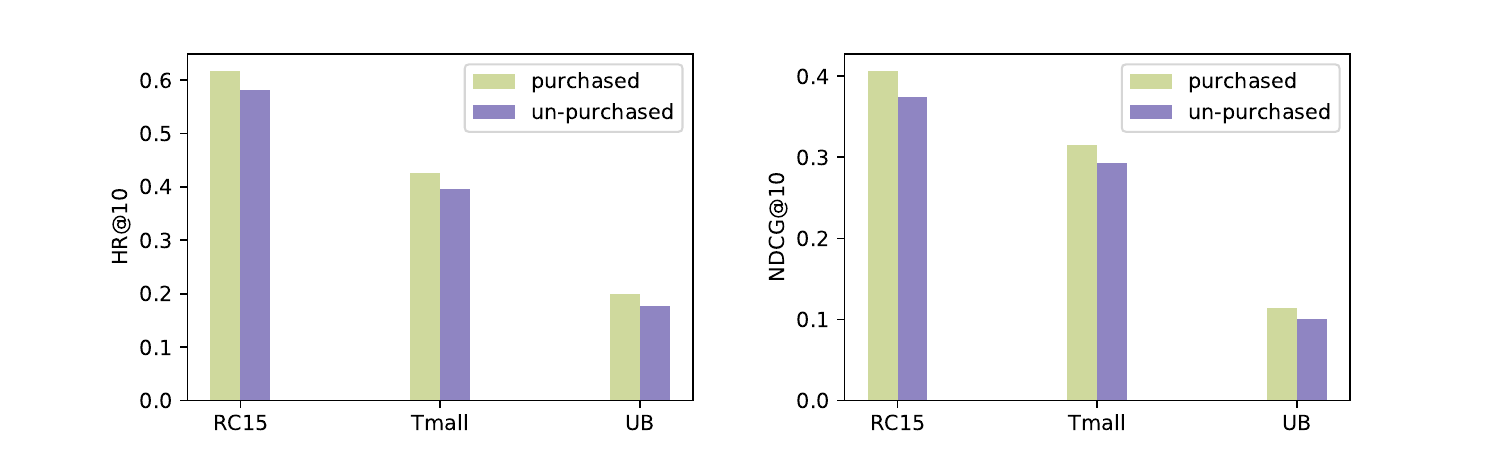}
\caption{ Recommendation performance in predicting purchased items versus un-purchased items on Rec15, Tmall, and UB. Notice that the behaviors in ML1M are simulated, which are thus not included.}
\label{fig:nonpurchase}
\end{figure}

\indent To investigate how the PIP module contributes to the {final purchased item prediction}, we conduct an empirical study on {the behavior types of} the predicted items. 
Specifically, we select the last purchased item {of each sequence} as the test set {in the previous experiments.} 
In contrast, we choose an auxiliary behavior between the last two purchase behaviors as the test set {in this study. To ensure fairness of the comparison experiment, we strictly maintain the integrity of the sequence.} 
ML1M is a simulated dataset without purchase and auxiliary behaviors. 
Therefore, we do not include ML1M in the study. 
From Figure~\ref{fig:nonpurchase}, we can see that the accuracy of predicting purchase behavior is higher than that of predicting auxiliary behavior. 
The PIP module can capture the user's purchase intent well, {leading} to performance improvement.

\begin{figure}
	\begin{center}
		\includegraphics[scale = 0.35]{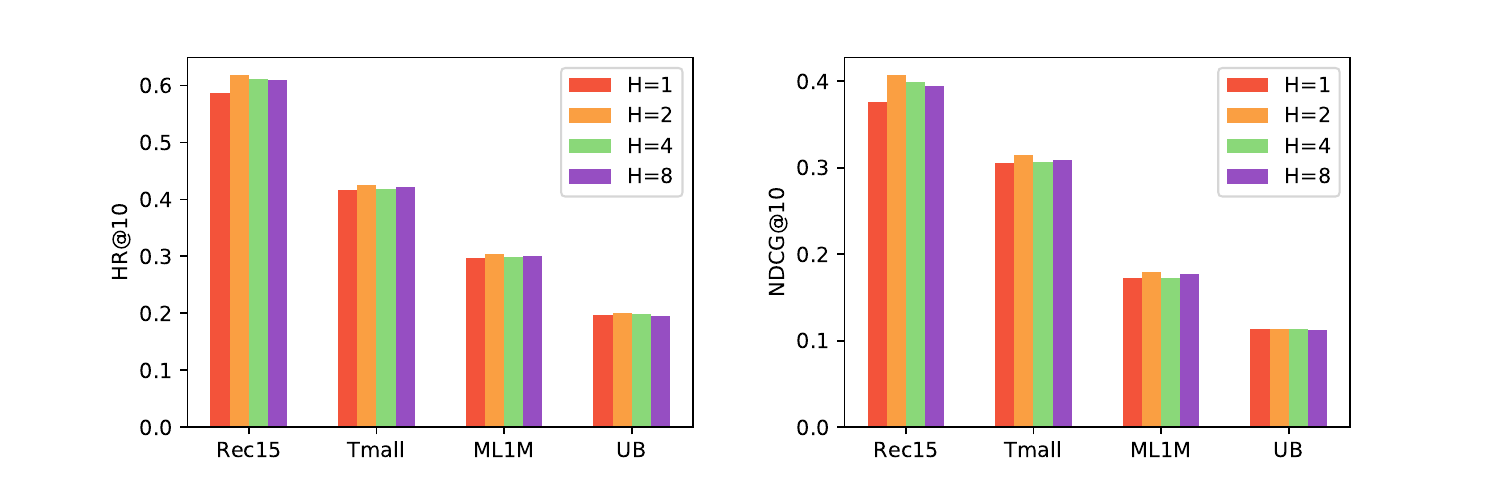}
	\end{center}
	\caption{Recommendation performance of our BMLP with different numbers of heads on four datasets.}
	\label{fig:head}
\end{figure}

\begin{figure}
	\begin{center}
		
		\small
		\begin{tabular}{cccc}
			\subfigure[Rec15]{\includegraphics[scale = 0.22]{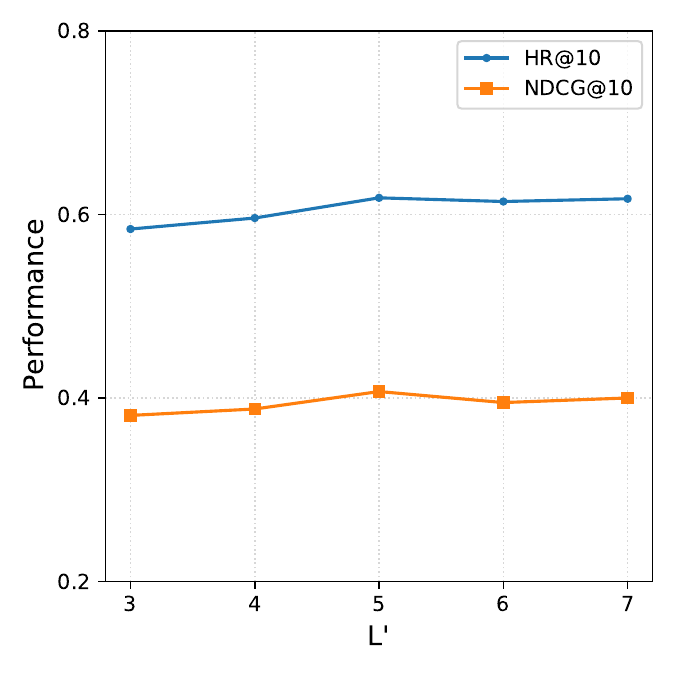}} &
			\subfigure[Tmall]{\includegraphics[scale = 0.22]{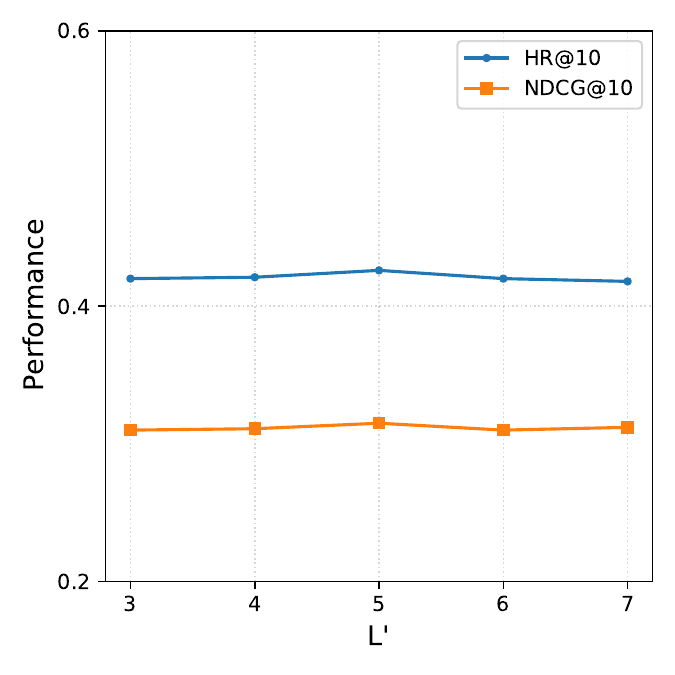}} &
			\subfigure[UB]{\includegraphics[scale = 0.22]{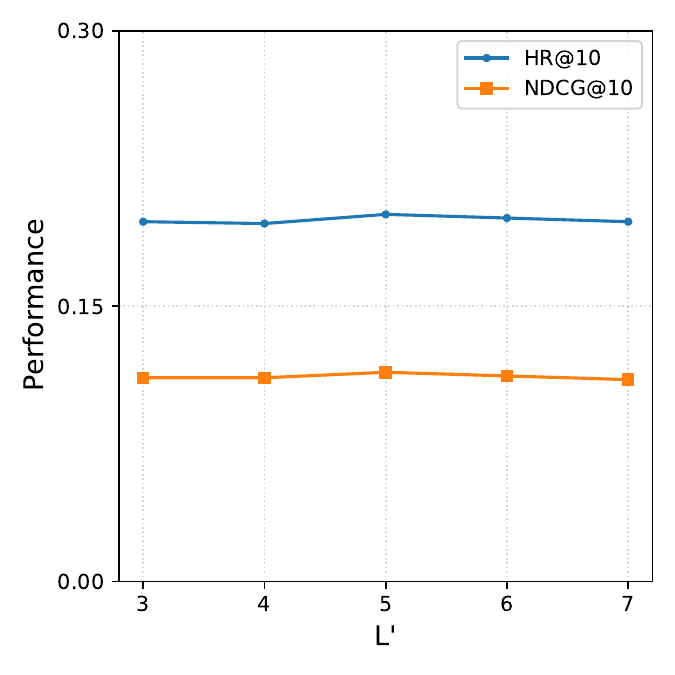}}
			
		\end{tabular}
		
	\end{center}
	
	\caption{Recommendation performance of our BMLP with different lengths of the most recent auxiliary behavior subsequences, i.e., {$L' \in \{3,4,5,6,7\}$}, on four datasets.}
	
	\label{fig:clilck_L_parameter}
\end{figure}
\begin{table*}[tp]
\caption{{The results of the grouped experiments on Tmall and UB. The term "examined" refers to the results of the test samples where the target item appears in the previous heterogeneous sequence, while "unexamined" indicates the contrary. The term "average" represents the overall results without distinguishing whether the target item appears in previous heterogeneous sequences.}}
	\label{tb:dup_nondup_result}
	\renewcommand\arraystretch{1.5}
	\begin{center}
		\setlength{\tabcolsep}{1.2mm}{
			\scriptsize
			\begin{tabular}{l | c c c c c c | c c c c c c}
				
				\hline\hline
				
				&  \multicolumn{6}{c|}{Tmall}	&\multicolumn{6}{c}{UB}\\ 
                
                &  \multicolumn{2}{c}{Examined}  & \multicolumn{2}{c}{Unexamined} &  \multicolumn{2}{c|}{Average}	 & \multicolumn{2}{c}{Examined}  & \multicolumn{2}{c}{Unexamined} &  \multicolumn{2}{c}{Average}\\ 
				
				&HR@10 & NDCG@10  &HR@10 & NDCG@10 
                &HR@10 & NDCG@10 &HR@10 & NDCG@10
                &HR@10 & NDCG@10 &HR@10 & NDCG@10\\
				
				\hline	
				\multirow{1}{*}{MBGMN}
				&0.0311 &	0.0185 &0.0278	&0.0168 & 0.0298 	& 0.0178	&0.0171 	&0.0079	&0.0152	&0.0066	&0.0162	&0.0073 \\

				\multirow{1}{*}{CML}
				
				&	0.0322&	0.0199& 0.0307&	0.0191 & 0.0316	&0.0196 &	0.0186	& 0.0092 & 0.0175	&0.0086	&0.0181	&0.0089
\\
            \multirow{1}{*}{MMCLR}
				
				&	0.0443&	0.0268& 0.0392&	0.0233 & 0.0423	&0.0254&	0.0261	& 0.0120 & 0.0211	&0.0103	&0.0237	&0.0112\\

    \multirow{1}{*}{SASRec}
				
				&	0.5575 &	0.4373& 0.0858&	0.0459 & 0.3688&0.2812&	0.2931	& 0.1737 & 0.0617	&0.0332	&0.1813	&0.1058\\

				\multirow{1}{*}{BMLP}
				
				& \bf{0.6483}
				& \bf{0.4889}
				&\bf 0.0916
				& \bf0.0523
                & \bf{0.4263}
				& \bf{0.3148}
				&\bf 0.3235
				& \bf 0.1844 & \bf 0.0687	&\bf 0.0391	&\bf 0.2004	&\bf 0.1142\\
				\hline\hline
			\end{tabular}
		}
	\end{center}
\end{table*}
{\subsection{Performance Comparison with Heterogeneous Non-sequential Recommendation Methods (RQ6)}
In this subsection, we compare several classic heterogeneous non-sequential recommendation methods and conduct a deep analysis of how the appearance or non-appearance of a target item in a historical sequence affects the performance. Firstly, we conduct an analysis on the percentage of target items appearing in recent historical behaviors on all datasets. Among all the test samples, the percentage of the target items appearing in the previous historical behaviors is 60.1\%, 51.7\%, 0\%, and 99.6\% on Tmall, UB, ML1M, and Rec15, respectively. The ML1M dataset is a simulated dataset that lacks purchase and click behaviors. In contrast, the Rec15 dataset comprises a significant number of instances where items are first clicked and subsequently purchased, leading to a remarkably high occurrence rate of 99.6\%. To ensure the rigorousness of our experiments, we divide the test data into two groups, i.e., an unexamined test set and an examined test set on Tmall and UB. Notice that "examined" means that the target item appears in a previous heterogeneous sequence, while "unexamined" denotes the contrary. We conduct an evaluation of our proposed BMLP, in comparison with three heterogeneous non-sequential recommendation methods, i.e., MBGMN~\cite{MBGMN}, CML~\cite{CML} and MMCLR~\cite{MMCLR} and a classical sequential recommendation method SASRec. The results are shown in Table~\ref{tb:dup_nondup_result}. According to the experimental results, we can obtain the following observations: 1) Two sequential recommendation methods demonstrate a slightly superior performance on the examined test set compared to the overall average, while they experience a substantial decline on the unexamined test set. It suggests that the presence of the target item in the historical sequences simplifies the prediction task to some extent. 2) Regarding non-sequential recommendation, the performance degradation on the unexamined test set is not that much, indicating that the non-sequential recommendation methods are not particularly sensitive to the occurrence of the target item in historical sequences. 3) When comparing the results on the unexamined test set, our proposed method BMLP, still outperforms both the heterogeneous non-sequential recommendation methods and the sequential recommendation method SASRec. Notice that a previous study \cite{23dyn} has also observed that heterogeneous non-sequential recommendation methods exhibit particularly poor performance.}
\subsection{Hyperparameter Sensitivity (RQ7)}
\indent In this subsection, we explore the impact of two hyperparameters on our model, {i.e.,} the number of heads and the lengths of the most recent auxiliary behavior subsequences.
\subsubsection{The impact of the number of heads} 
We conduct the corresponding experiments with $H\in \{1, 2, 4, 8\}$. 
The experimental results on four datasets are shown in Fig~\ref{fig:head}. 
We find that the overall performance of our model reaches the best when $H$ $=2$. 
When $H$ is too large or too small, the performance is slightly worse. 
A larger value of $H$ means that the feature dimension on each head {is} correspondingly smaller.
If the feature dimension is too small, the information of the subspace cannot be {accurately} represented. 
A moderate value of $H$ can balance the dimensionality of each subspace and the number of subspaces.
\subsubsection{The impact of the length of the auxiliary behavior subsequences} 
The recent auxiliary behavior subsequences are used as the input of the PIP module, of which the length is important for {the} dissection of the PIP module. 
Therefore, we conduct comparison experiments with different lengths of these subsequences on each dataset, and the values of the length are selected from $\{3, 4, 5, 6, 7\}$. 
The experimental results are shown in Figure~\ref{fig:clilck_L_parameter}. 
We can see that the overall performance of our model reaches the best when $L'=5$. 
When the lengths of the most recent auxiliary behavior subsequences are relatively large or small, the model performance will decrease. 
The previous auxiliary behaviors become less important for the next purchased item when $L'$ is too large. 
Moreover, when the length is too small, it may introduce noise to the short-term purchase interest learned by the PIP module.

\section{CONCLUSIONS AND FUTURE WORK}
\indent In this paper, we propose a novel pure MLP-based solution, i.e., behavior-aware MLP (BMLP), for heterogeneous sequential recommendation (HSR). 
Specifically, {our} BMLP contains three modules, i.e., heterogeneous interest perception (HIP), purchase intent perception (PIP), and recommendation. 
The main structure of HIP and PIP {consists} of two MLPs, one dealing with sequential information and the other with feature representation.
In the HIP module, we consider behavior types and behavior transition relations for modeling behaviors at multiple granularities to capture a user's heterogeneous interests. 
In the PIP module, we adaptively aggregate the recent auxiliary behavior subsequences to obtain a user's dynamic purchase intent. 
Finally, the recommendation module fuses the heterogeneous interest and the purchase intent for prediction. 
Extensive experiments show that our BMLP outperforms the state-of-the-art baselines.
\\
\indent {For future works, we are interested in leveraging rich contextual information, such as review comments and item attributes. Moreover, we intend to generalize our model to multi-task scenarios in order to improve the performance on both the purchase and non-purchase behaviors. }

\Acknowledgements{We thank the support of National Natural Science Foundation of China No. 62172283 and No. 62272315. We thank Miss Qianzhen Rao for her helpful discussions.}





\par\noindent 
\parbox[t]{\linewidth}{
\noindent\parpic{\includegraphics[height=1.5in,width=1in,clip,keepaspectratio]{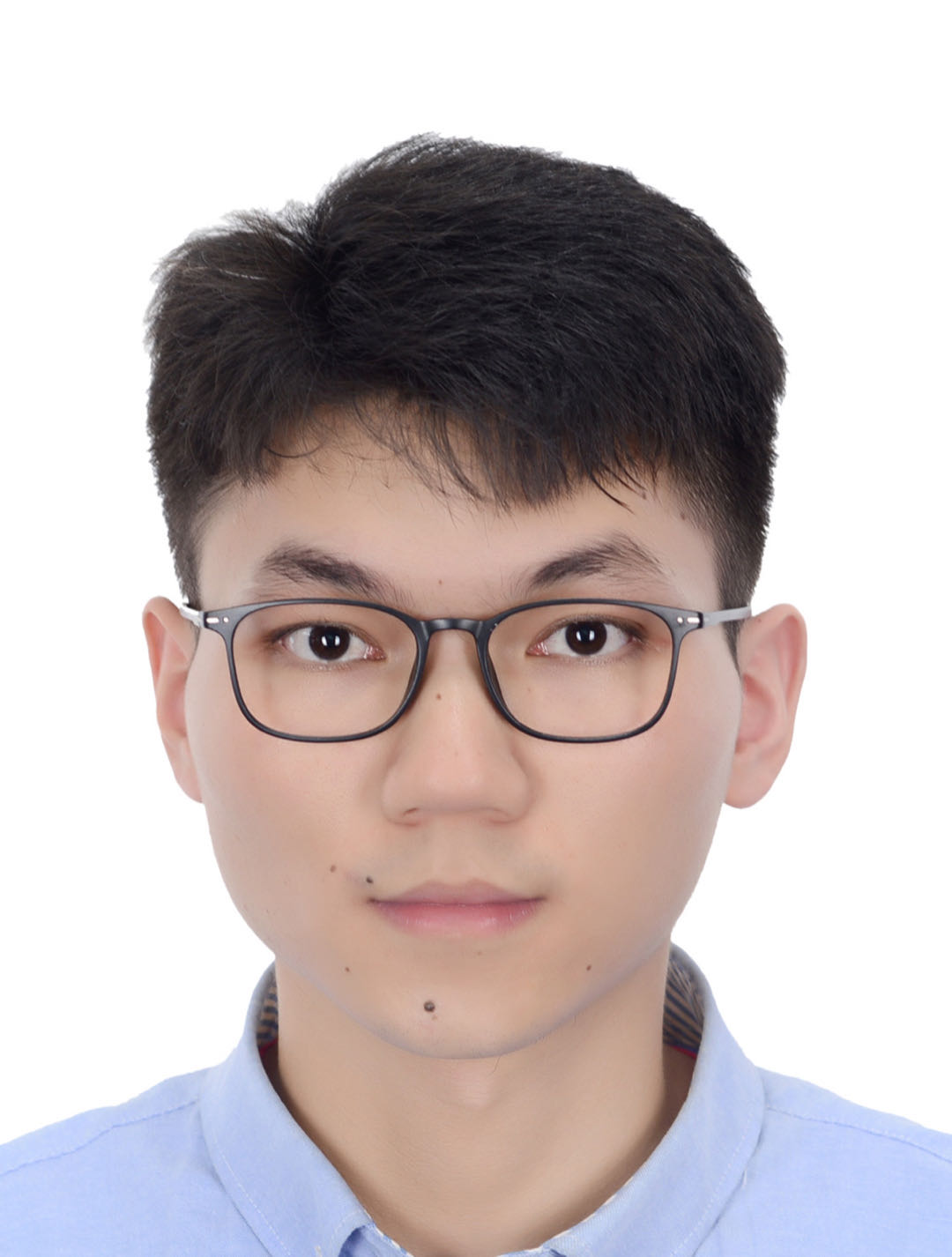}}
\noindent {\bf Weixin Li} is currently pursuing the M.S. degree with the College of Computer Science and Software Engineering, Shenzhen University, China. His research interests include recommender systems and deep learning. Contact him at liweixin2021@email.szu.edu.cn.}
\vspace{2\baselineskip}


\par\noindent 
\parbox[t]{\linewidth}{
\noindent\parpic{\includegraphics[height=1.5in,width=1in,clip,keepaspectratio]{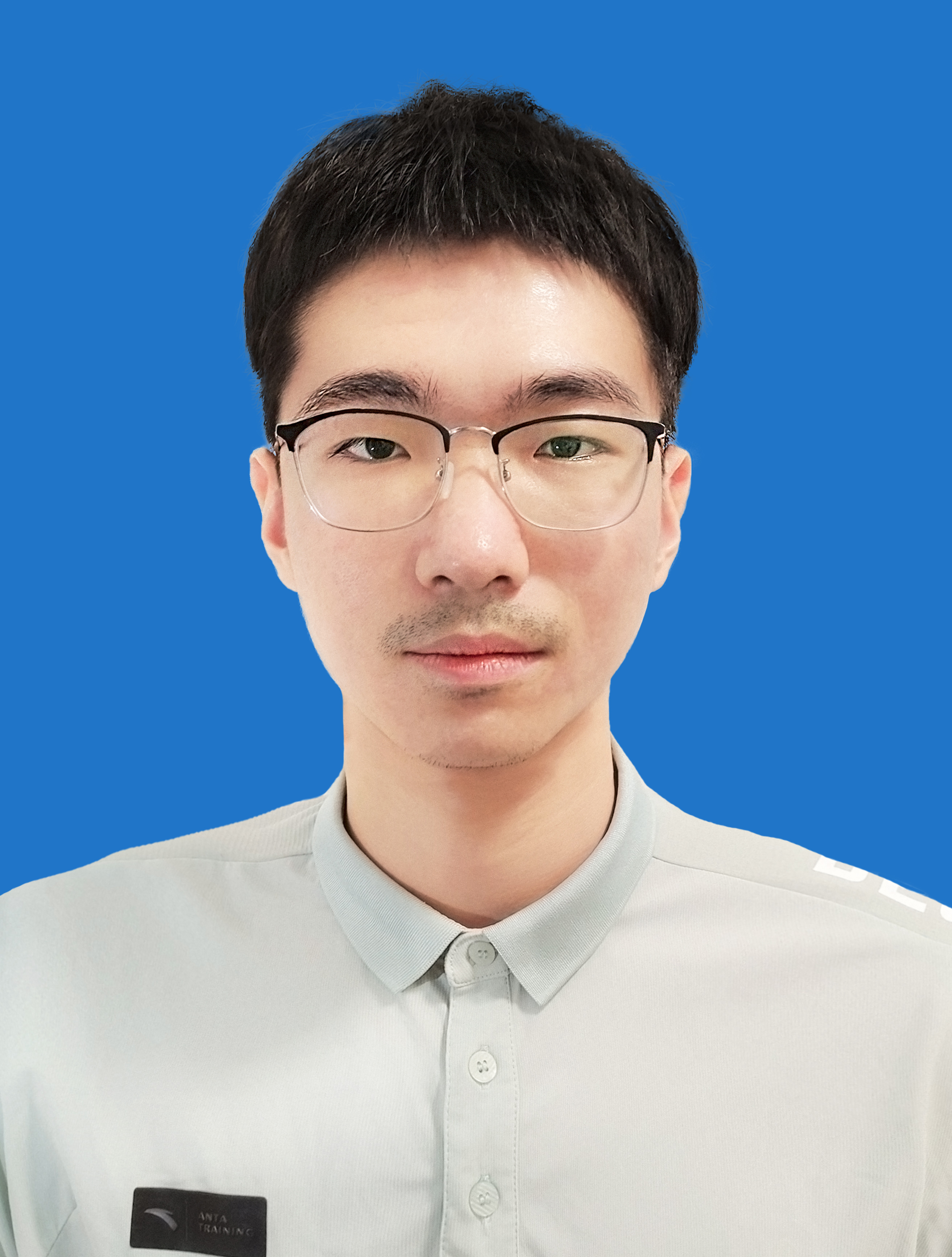}}
\noindent {\bf Yuhao Wu} is currently pursuing the Ph.D. degree with the College of Computer Science and Software Engineering, Shenzhen University, China. His research interests include zero-shot learning, recommender systems and machine learning. He has published papers in TITS, DMDK, KSEM, ICA3PP, ICMLC, RecSys and Neurocomputing. Contact him at wuyuhao2020@email.szu.edu.cn}
\vspace{2\baselineskip}

\par\noindent 
\parbox[t]{\linewidth}{
\noindent\parpic{\includegraphics[height=1.5in,width=1in,clip,keepaspectratio]{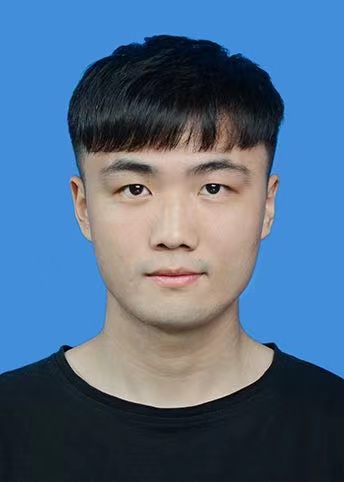}}
\noindent {\bf Yang Liu} is currently pursuing the M.S. degree with the College of Computer Science and Software Engineering, Shenzhen University, China. His research interests include recommender systems and deep learning. He has published papers in RecSys and ICDM. Contact him at 2110276162@email.szu.edu.cn.
}
\vspace{2\baselineskip}

\par\noindent 
\parbox[t]{\linewidth}{
\noindent\parpic{\includegraphics[height=1.5in,width=1in,clip,keepaspectratio]{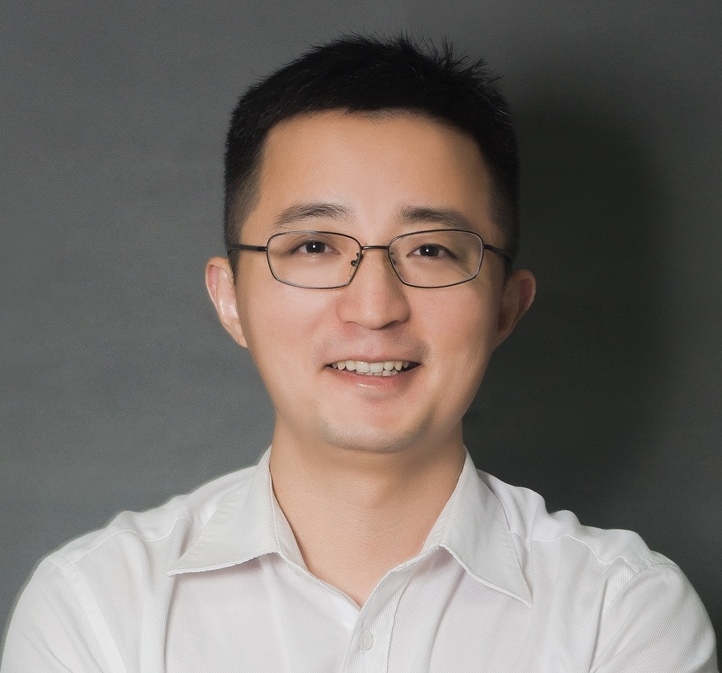}}
\noindent {\bf Weike Pan} received the Ph.D. degree in Computer Science and Engineering from the Hong Kong University of Science and Technology, Kowloon, Hong Kong, China, in 2012. He is currently a professor with the College of Computer Science and Software Engineering, Shenzhen University, Shenzhen, China. His research interests include transfer learning, federated learning, recommender systems and machine learning. He has published research papers in AIJ, TBD, TIIS, TIST, TKDE, TOIS, AAAI, CIKM, IJCAI, RecSys, SDM, SIGIR, WSDM, etc. Contact him at panweike@szu.edu.cn.
}
\vspace{2\baselineskip}

\par\noindent 
\parbox[t]{\linewidth}{
\noindent\parpic{\includegraphics[height=1.5in,width=1in,clip,keepaspectratio]{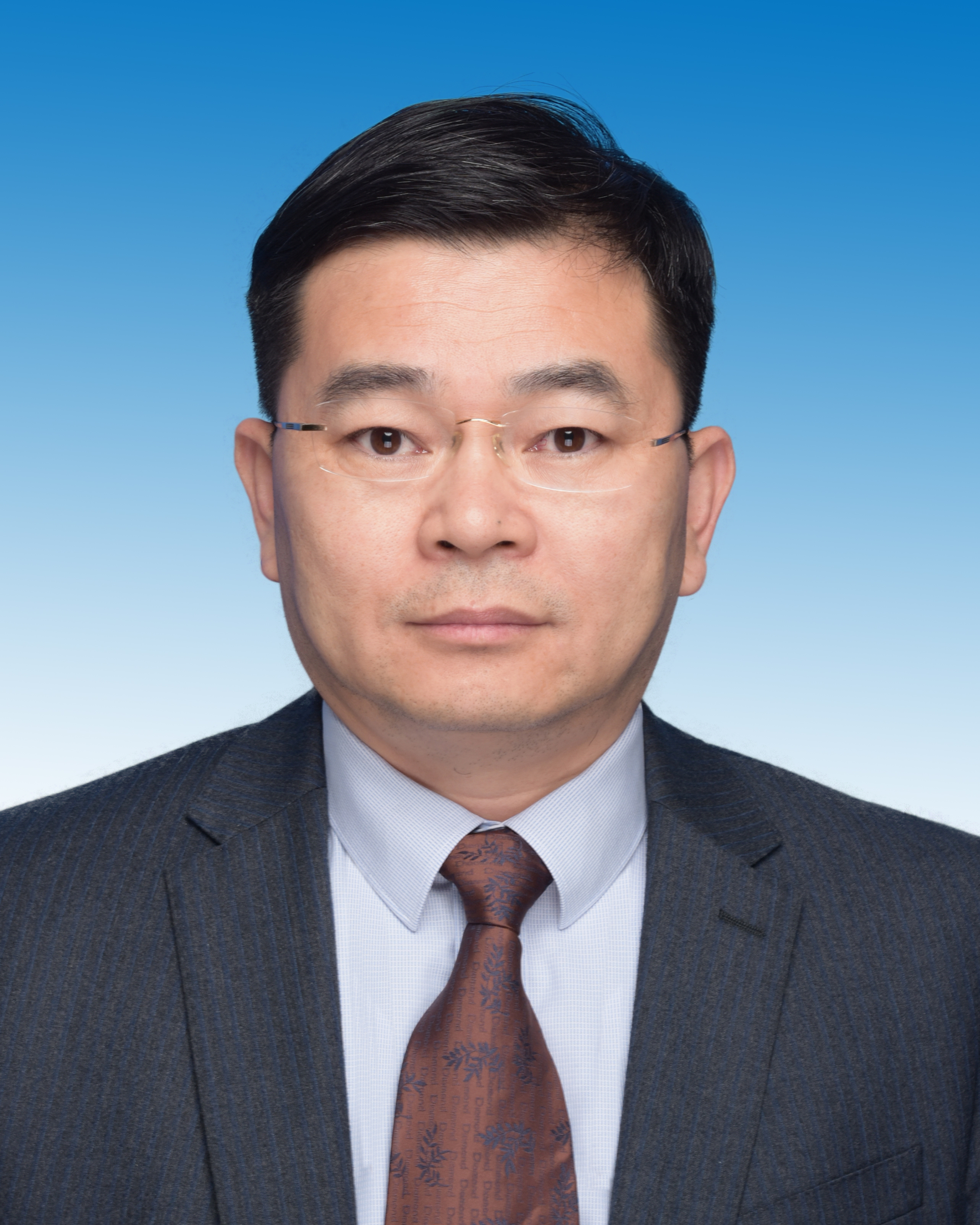}}
\noindent {\bf Zhong Ming} received the Ph.D. degree in Computer Science and Technology from the Sun Yat-Sen University, Guangzhou, China, in 2003. He is currently a professor with the College of Computer Science and Software Engineering, Shenzhen University, Shenzhen, China. His research interests include software engineering and artificial intelligence. He has published more than 200 refereed international conference and journal papers (including 40+ ACM/IEEE Transactions papers). He was the recipient of the ACM TiiS 2016 Best Paper Award and some other best paper awards. Contact him at mingz@szu.edu.cn.}
\vspace{2\baselineskip}

\end{document}